\documentstyle[12pt]{article}
\newcommand{\be}{\begin{equation}}
\newcommand{\ee}{\end{equation}}

{
\begin{document}

\begin{center}
{\large\bf Unitary relations in time-dependent harmonic oscillators }
\end{center}
\begin{center}
Dae-Yup Song\footnote[2]{E-mail address: 
        dsong@sunchon.sunchon.ac.kr}
\end{center}
\begin{center}
{ Department of Physics,\\ Sunchon National University, Sunchon 
540-742, Korea}
\end{center}
\bigskip
\begin{center}
Short title: The time-dependent harmonic oscillators \\
Classification numbers: 03.65.Fd  03.65.Ca
\end{center}
\bigskip
\begin{abstract}
For a harmonic oscillator with time-dependent (positive) mass and frequency, 
an unitary operator is shown to transform the quantum states of the system 
to those of a harmonic oscillator system of unit mass and time-dependent 
frequency, as well as operators. For a driven harmonic oscillator,
an unitary transformation which relates the driven system and the system 
of same mass and frequency without driving force is given, as a generalization
of previous results, in terms of the solution of classical equation 
of motion of the driven system.
These transformations, thus, give a simple way of finding exact wave
functions of a driven harmonic oscillator system, provided the quantum 
states of the corresponding system of unit mass are given. 

\end{abstract}
\newpage
\section{Introduction}
The harmonic oscillators with time-dependent mass and 
frequency have long been of interest and give examples of
exactly solvable time-dependent systems. For the oscillator of constant 
mass and time-dependent frequency, Lewis \cite{Lewis,LR} has shown 
that there exists quantum mechanically invariant operator, unaware of 
Ermakov's results \cite{Erma}. This so-called Ermakov-Lewis invariant 
operator can be used to find exact quantum states. This method 
has then been generalized to include time-dependent mass \cite{Ji,Yeon},
driving force \cite{Kim}, and to a general quadratic system whose 
Hamiltonian has all terms of position and momentum quadratic or less
than that \cite{Lee,Yeon1}.

Another systematic method to find exact quantum states of the systems 
is to use the Lagrangian formulation of Feynman and Hibbs \cite{FH} who
have shown that the position-dependent part of the kernel (propagator)
is determined from classical action. This observation by Feynman and 
Hibbs gives a good explanation of the fact that the wave functions of 
the quantum states are described in terms of solutions of classical 
equation of motion. In \cite{Song}, this method has been developed
to give the exact kernel. The wave functions of general quadratic 
systems are then found by factorizing the kernel.

With these generalizations from the Lewis's results, one important 
question arises: Do the generalizations give quite new systems?
This question has long been studied through the canonical 
transformation in classical mechanics \cite{Leach,Ped}. 
In quantum treatment \cite{Li,Mosta}, in addition to the 
recognition of relation between driven system and undriven 
system \cite{Husimi,Haar}, a part of answer to this 
question has been given by Mostafazadeh \cite{Mosta}. 
He has found an unitary operator which transforms the 
Hamiltonian of the oscillator of time-dependent mass and frequency to 
that of constant mass. So, one of his conclusions is the confirmation, 
in quantum treatment, of that the old (classical) result that
Hamiltonian of the Caldirola-Kanai (C-K) system \cite{CK,Kanai} can be 
obtained from that of a simple harmonic oscillator \cite{Kerner}.

The purpose of this paper is to show that the generalizations
\cite{Ji,Yeon,Kim,Lee,Yeon1,Song}
of Lewis's results can be done through the unitary transformation 
not only in operator level but also in representation theory.
For this, we need two unitary transformations.
One of the transformations is to relate driven harmonic oscillator 
system to that of the same parameters without driving force. 
The operator of this transformation will be given in terms of solution 
of classical equation of motion of the driven system, as a 
generalization of previous results \cite{Husimi,Haar,Kerner}. 
The other transformation is to 
change the mass and frequency of the system. The mass-frequency
relation given by Mostafazadeh \cite{Mosta} will be obtained 
also by comparing the classical equations of motion of the two systems. 
If we choose proper 
parameters which will be explicitly found, the transformation changes 
the system of time-dependent (positive) mass and frequency to 
that of unit mass. 

By applying the operators to the quantum states of the system of 
unit mass, it will be shown that the wave functions of driven 
harmonic oscillator can be obtained from those of the 
corresponding undriven system of unit mass.
Therefore, this transformation method gives a simple way of 
finding exact quantum states 
of a driven harmonic oscillator system \cite{Kim}
or a general quadratic system \cite{Lee,Yeon1,Song}, provided quantum 
states of the corresponding system of unit mass are given. 
As explicit examples, we consider two models which are equivalent 
to simple harmonic oscillators. One of them is the C-K system \cite{CK,Kanai} 
and the wave functions of this system will be evaluated 
from those of simple harmonic oscillators.

\section{The unitary transformations for harmonic oscillator systems
without driving force}
We start with the transformation for the time-dependent Hamiltonian
\be
H(p,x,t)= {p^2 \over 2 M(t)} +{1\over 2} M(t)w^2(t)x^2,
\ee
where $M(t)$ and $w(t)$ are time-dependent (positive) mass and  frequency, 
respectively. 
Then the wave function $\psi(x,t)$ of a quantum eigenstate should satisfy 
the Schr\"{o}dinger equation
\be
O\psi(x,t)=0,~~ {\rm with} ~~
O\equiv-i\hbar {\partial\over \partial t} 
          +H({\hbar\over i}{\partial\over \partial x},x,t).
\ee
Since we will consider the time-dependent unitary transformation, 
it is necessary to consider transformation of the operator $O$ 
instead of $H$ \cite{Li,Mosta,Uni}. 
With the unitary operator, $U_c$, defined as
\be
U_c= e^{i\alpha x^2 /\hbar}e^{i\beta(xp+px)/4\hbar},
\ee
one may find the relation
\be
U_c OU_c^\dagger=-i\hbar{\partial\over \partial t}
      +{p^2\over 2Me^\beta} 
      + (xp+px)[-{\dot{\beta}\over 4}-{\alpha\over Me^\beta}]  
      +{x^2\over 2}[Mw^2e^\beta +2\alpha\dot{\beta}-2\dot{\alpha}
               +{4\alpha^2 \over Me^\beta}],
\ee
where the dots over variables denote the differentiation 
with respect to time. Equation (4) implies that the unitary 
transformation gives rise to a new system described by the 
Hamiltonian 
\be
H_{new}={p^2\over 2Me^\beta} 
      + (xp+px)[-{\dot{\beta}\over 4}-{\alpha\over Me^\beta}]  
      +{x^2\over 2}[Mw^2e^\beta +2\alpha\dot{\beta}-2\dot{\alpha}
               +{4\alpha^2 \over Me^\beta}].
\ee
As is well-known, the term proportional to $(xp+px)$ in Hamiltonian
can be generated by acting unitary transformation in Hamiltonian 
formulation \cite{Mosta2}, or by adding the term proportional 
to $dx^2/dt$ to the Lagrangian \cite{Song}. Since the term 
proportional to $(xp+px)$ can be interpreted as a result of 
simple unitary transformation, we will take $\alpha$ as
\be
\alpha= -{M\over 4} \dot{\beta} e^\beta.
\ee 
With this relation, $H_{new}$ is written as 
\be
H_{new}={p^2\over 2Me^\beta} 
        +Me^\beta[w^2+{1\over 2}{\dot{M}\over M}\dot{\beta}
           +{\ddot{\beta}\over 2} +{\dot{\beta}^2\over4}]
              {x^2\over 2}.
\ee
The $H_{new}$ in equation (7) shows \cite{Mosta,Uni} that unitary 
transformation can be used to find a new harmonic oscillator system which 
has different mass and frequency from the original system of equation 
(1). Among these systems, we can find a system of unit mass by taking
\be
\beta= -\ln M(t),
\ee
which is described by the Hamiltonian
\be
H_0={p^2 \over 2 } 
    + {1\over 2} (w^2+{1\over 4} ({\dot{M}\over M})^2
         -{1\over 2}{\ddot{M}\over M})x^2
   ={p^2 \over 2 } + {1\over 2} 
        (w^2 -{1\over \sqrt{M}}{d^2\sqrt{M}\over dt^2})x^2.
\ee  
That is, the mass of the system is 1, while the new frequency, $w_0$,
is given by \cite{Mosta}
\be
w_0^2(t)=w^2 -{1\over \sqrt{M}}{d^2\sqrt{M}\over dt^2}.
\ee
The unitary operator for the transformation from the Hamiltonian 
in equation (1) to $H_0$ is now given as
\be
U_0=\exp({i\over 4\hbar}{\dot{M}\over M}x^2)
      \exp(-{i\ln M\over 4\hbar}(xp+px)). 
\ee
In the above equations, unit mass which has not been written explicitly 
should be taken into account to find the correct physical dimensions,
which will also be true from now on.

One may find that the unitary operator in equation (11) \cite{Mosta}
which does not depend on the solutions of the classical equation of motion 
is different from that in \cite{Li}.

The system described by Hamiltonian in equation (9) is one of those 
considered by Lewis \cite{Lewis}. With non-negative integer $n$,
the $n$-order Hermite polynomial $H_n$ and two linearly independent 
real solutions $u_0(t),~v_0(t)$ of classical equation of motion
\be
\ddot{\bar{x}_0} +  w_0^2(t) \bar{x}_0 =0,
\ee
the wave functions of quantum eigenstates are given as 
\cite{Lewis,Ji, Yeon,Song}  
\begin{eqnarray}
\psi_n^0(x,t)&=& 
     {1\over \sqrt{2^n n!}}({\Omega_0 \over \pi\hbar})^{1\over 4}
     {1\over \sqrt{\rho_0(t)}}[{u_0(t)-iv_0(t) \over \rho_0(t)}]^{n+{1\over 2}}
     \exp[{x^2\over 2\hbar}(-{\Omega_0 \over \rho_0^2(t)}
               +i {\dot{\rho}_0(t) \over \rho_0(t)})] 
\cr&&~~~~~~~~~~~~
      \times    H_n(\sqrt{\Omega_0 \over \hbar} {x \over \rho_0(t)}).
\end{eqnarray}
In equation (13), $\Omega_0, \rho_0(t)$ are defined as
\be
\Omega_0=[\dot{v}_0(t)u_0(t) - \dot{u}_0(t)v_0(t)],~~~~
\rho_0(t)=\sqrt{u_0^2(t) +v_0^2 (t)}.
\ee
$\Omega_0$ which depends on the choice of classical 
solutions is constant along time evolution. Even though the corresponding 
Schr\"{o}dinger equation is formally satisfied for any non-zero $\Omega_0$, 
we will only consider the cases of positive $\Omega_0$ for applications.

For the simple harmonic oscillator of unit mass and positive constant 
frequency $w_s$, one may take the classical solutions as
$u_0=A\cos w_st,$ and $v=B\sin w_st$, with positive constants $A$ and $B$.
The wave functions in equation (13) then becomes 
\begin{eqnarray}
\psi_n^{SHO}(w_s;x,t)&=& 
     {1\over \sqrt{2^n n!}}({Cw_s \over \pi\hbar})^{1\over 4}
     {1\over \sqrt{\tilde{\rho}_s(t)}}
     [{C\cos w_st-i\sin w_st \over 
                   \tilde{\rho}_s(t)}]^{n+{1\over 2}}
  \cr
  & &~~~~~~~~~~~~~\times
     \exp({x^2\over 2\hbar}[-{Cw_s \over \tilde{\rho}_s^2(t)}
               +i {\dot{\tilde{\rho}}_s(t) \over \tilde{\rho}_s(t)}])
    H_n(\sqrt{Cw_s \over \hbar} 
                      {x \over \tilde{\rho}_s(t)}),
\end{eqnarray}
where 
\be
\tilde{\rho}_s(w_0)=\sqrt{1+(C^2-1)\cos^2 w_0t}~~~{\rm and}~~~
C= {A\over B}.
\ee
With the choice of $C=1$, $\psi_n^{SHO}(w_s, x,t)$ reduces 
to the usual stationary wave function of the unit mass simple
harmonic oscillator; However, for $C\neq 1$,
the wave functions describe the quantum eigenstates of pulsating 
probability distribution.

The unitary transformation changes quantum states as well as 
operators. For showing this fact explicitly, we define a set of two 
linearly independent functions $\{u,v\}$ as 
\be
u(t)={u_0(t) \over \sqrt{M}},~~~v(t)={v_0(t) \over \sqrt{M}}.
\ee
One then easily find that $\{u,v\}$ satisfies
the differential equation
\be
{d \over {dt}} (M \dot{\bar{x}}) + M(t) w^2(t) \bar{x} =0 
~~~{\rm or}~~~
\ddot{\bar{x}} +{\dot{M} \over M}\dot{\bar{x}} + w^2(t) \bar{x} =0, 
\ee
which is the classical equation of motion for the system described
by the Hamiltonian in equation (1).
Furthermore, by substituting $\bar{x}$ with $\bar{x}_0/\sqrt{M}$
in equation (18) and comparing the equations (12,18), one may 
{\it reproduce} the mass-frequency relation (10).
We also define $\Omega, \rho(t)$ as
\be
\Omega=M(t)[\dot{v}(t)u(t) - \dot{u}(t)v(t)],~~~~
\rho(t)=\sqrt{u^2(t) +v^2 (t)}.
\ee
$\Omega$ is then constant along time.
Making use of the fact that
\be
e^{(a(t)x{\partial\over \partial x})} f(x)= f(e^{a(t)} x),
\ee
{\it through the unitary transformation},  
one may find the wave function for the system of the Hamiltonian
in equation (1):
\begin{eqnarray}
\psi_n(x,t)&=& U_0^\dagger\psi_n^0\\
&=& 
     {1\over \sqrt{2^n n!}}({\Omega \over \pi\hbar})^{1\over 4}
     {1\over \sqrt{\rho(t)}}[{u(t)-iv(t) \over \rho(t)}]^{n+{1\over 2}}
     \exp[{x^2\over 2\hbar}(-{\Omega \over \rho^2(t)}
               +i M(t){\dot{\rho}(t) \over \rho(t)})]
\cr & &~~~~~~~~
          H_n(\sqrt{\Omega \over \hbar} {x \over \rho(t)}) 
\end{eqnarray}
which agrees with the known result \cite{Ji,Yeon,Song}.

\section{Examples}
We consider two systems which are unitarily equivalent to the 
simple harmonic oscillator, as examples.
The first one is the C-K system \cite{CK,Kanai} described by the 
Hamiltonian:
\be
H^{C-K}(p,x,t)= {p^2 \over 2 m e^{\gamma t}} +{1\over 2} me^{\gamma t}w_1^2x^2,
\ee 
with constant $m,~\gamma,$ and $w_1$. Equation (10) shows  that
the C-K system is unitarily equivalent to the simple harmonic oscillator
of unit mass and constant frequency $w_{ck}$, where $w_{ck}$ is given by
\cite{Kerner}
\be
w_{ck}^2=w_1^2 -{\gamma^2 \over 4}.
\ee
For the case of positive real $w_{ck}$, the wave functions are easily found
from those in equation (15) by applying the relation in (21);
\begin{eqnarray}
\psi_n^{C-K}
&=&\exp({i\over 4\hbar}(\gamma t +\ln m)(xp+px))\exp(-{i\gamma \over 4\hbar}x^2)
             \psi_n^{SHO}(w_{ck};x,t)      \cr
&=&(me^{\gamma t})^{1\over 4}
     \exp({\gamma t+\ln m\over 2}x {\partial\over \partial x})
     \exp(-{i\gamma \over 4\hbar}x^2)\psi_n^{SHO}(w_{ck};x,t)      \cr
&=& {1\over \sqrt{2^n n!}}({me^{\gamma t}Cw_{ck} \over \pi\hbar})^{1\over 4}
     {1\over \sqrt{\tilde{\rho}_{ck}}}
     [{C\cos w_{ck}t-i\sin w_{ck}t \over 
                   \tilde{\rho}_{ck}}]^{n+{1\over 2}}
\cr &&
   \times  \exp[{me^{\gamma t}x^2\over 2\hbar}(-{Cw_{ck} \over \tilde{\rho}_{ck}^2}
       +i ({\dot{\tilde{\rho}}_{ck} \over \tilde{\rho}_{ck}}-{\gamma\over 2}))]
    H_n(\sqrt{me^{\gamma t}Cw_{ck} \over \hbar} 
                      {x \over \tilde{\rho}_{ck}}),
\end{eqnarray}
where 
\be
\tilde{\rho}_{ck}=\tilde{\rho}_s(w_{ck}).
\ee
By adjusting the $C$, the wave functions in equation (25) 
can be shown to give those in \cite{Hasse,Dod,CKE,PR}.  
By taking two linearly independent solution of the classical equation of motion: 
\[
\ddot{\bar{x}} +\gamma\dot{\bar{x}} + w^2(t) \bar{x} =0 
\]
of the C-K system as $u=Ae^{-{\gamma t\over 2}}\cos w_{ck}t$ and 
$v=Be^{-{\gamma t\over 2}}\sin w_{ck}t$, one can also obtain the 
wave functions in equation (25) from the formula (22).

As another example, we consider the system of the damped pulsating oscillator 
considered in \cite{Lo,Kim}, where the time dependent mass $M_{Lo}$ is given as
$M_{Lo}=m_0\exp[2(\gamma t+\mu \sin\nu t)]$ with constant $m_0,~\gamma,~\mu$ and 
$\nu$. The frequency $w(t)$ of the model is defined as 
$w^2=w_{Lo}^2+{1\over \sqrt{M_{Lo}}}{d^2 \sqrt{M_{Lo}} \over dt^2}$, 
with constant $w_{Lo}$. Though this model looks complicated, equation (10) implies 
that this system is unitarily equivalent to the simple harmonic oscillator of
unit mass and constant frequency $w_{Lo}$. The wave functions can also be obtained 
from those in equation (15) as
\begin{eqnarray}
\psi_n^{Lo}
&=& {1\over \sqrt{2^n n!}}({M_{Lo}Cw_{Lo} \over \pi\hbar})^{1\over 4}
     {1\over \sqrt{\tilde{\rho}_{Lo}}}
     [{C\cos w_{Lo}t-i\sin w_{Lo}t \over 
                   \tilde{\rho}_{Lo}}]^{n+{1\over 2}}
\cr &&
   \times  \exp[{M_{Lo}\over 2\hbar}x^2(-{Cw_{Lo} \over \tilde{\rho}_{Lo}^2}
       +i ({\dot{\tilde{\rho}}_{Lo} \over \tilde{\rho}_{Lo}}
            -{1\over 2}{\dot{M}_{Lo}\over M_{Lo}}))]
    H_n(\sqrt{M_{Lo}Cw_{Lo} \over \hbar} 
                      {x \over \tilde{\rho}_{Lo}}),
\end{eqnarray}
where 
\be
\tilde{\rho}_{ck}=\tilde{\rho}_s(w_{Lo}).
\ee

\section{The transformations for driven oscillator systems}
The driven harmonic oscillator is described by the Hamiltonian
\be
H^F= {p^2 \over 2 M(t)} +{1\over 2} M(t)w^2(t)x^2-xF(t).
\ee
To find the unitary transformation, we define the $x_p$ as a particular 
solution of the classical equation of motion:
\be
{d \over {dt}} (M \dot{x_p}) + M(t) w^2(t) x_p=F(t).
\ee
We also introduce a function $\delta(t)$ defined as
\be
\dot{\delta}= {Mw^2\over 2} x_p^2 -{M \over 2}\dot{x}_p^2.
\ee
By defining an operator $O_F$ as 
\be
O_F=-i\hbar {\partial \over \partial t} + H_F,
\ee  
making use of the equations (30,31), one can find the relation:
\be
U_FOU_F^\dagger=O_F,
\ee
where $U_F$ is given as
\be
U_F=\exp[{i\over \hbar}(M\dot{x}_px+\delta(t))]
          \exp(-{i\over \hbar}x_p p).
\ee 
The wave function for the system of the Hamiltonian in equation (29)
can thus be evaluated through the unitary transformation as
\begin{eqnarray}
\psi_n^F &=& U_F\psi_n\\
&=& U_FU_0^\dagger \psi_n^0 \\
&=&
     {1\over \sqrt{2^n n!}}({\Omega \over \pi\hbar})^{1\over 4}
     {1\over \sqrt{\rho(t)}}[{u(t)-iv(t) \over \rho(t)}]^{n+{1\over 2}}
     \exp[{i\over \hbar}(M\dot{x}_px+\delta(t))]
\cr & &~~~~~~~~
     \exp[{(x-x_p)^2\over 2\hbar}(-{\Omega \over \rho^2(t)}
               +i M(t){\dot{\rho}(t) \over \rho(t)})]
          H_n(\sqrt{\Omega \over \hbar} {x -x_p \over \rho(t)}). 
\end{eqnarray}
One can explicitly check that $\psi_n^F$ satisfy the Schr\"{o}dinger equation
\be 
O_F \psi_n^F=0 ~~~{\rm or}~~~ 
i\hbar{\partial \psi_n^F\over \partial t}=
  -{\hbar^2 \over 2M}{\partial^2 \over \partial x^2}\psi_n^F 
  +{Mw^2 \over 2}x^2 \psi_n^F -xF(t)\psi_n^F.
\ee

Through a different approach, the relation (35) has long been recognized 
as in \cite{Husimi,Haar} for special cases.

In \cite{Song} the wave functions for the driven harmonic oscillator are found
by factorizing the kernel. If $\delta$ is given as   
\be
\delta=-{M\over 2}{\dot{v}\over v}x_p^2-{1\over 2}
        \int_{t_0}^t M(z)(x_p(z){\dot{v}\over v}-\dot{x}_p(z))^2dz
\ee
with an arbitrary constant $t_0$, the wave functions in equation (37) 
reduce to those in \cite{Song}. And one may easily check that the 
$\delta(t)$ in equation (39) satisfies the relation (31). The defining
relation (31), however, suggests a simpler form  $\delta(t)$ as 
\be
\delta(t)=\int_{t_0}^t[{M(z)w^2(z)\over 2} x_p^2(z) 
                -{M(z) \over 2}\dot{x}_p^2(z) ] dz,
\ee
which can be shown  equal to that in equation (39), up to a constant,
by making use of the equation of motion in (18). 

For a given particular solution $x_p(t)$, new solutions can be obtained
by adding linear combinations of homogeneous solutions. For instance,
a new solution $x_p'(t)$ can be given as $x_p(t)+Cu(t)$. The $\delta(t)$
depend on the choice of the classical solution, and the difference of 
$\delta$ evaluated with $x_p'(t)$ from that with $x_p(t)$ is written as
$-CM\dot{u}(x_p+{1 \over 2}Cu)$ up to a additive constant.

\section{Summary and discussions}
In summary we have found the unitary relations between the systems of 
time-dependent harmonic oscillators. The first relation is between the systems of
time-dependent mass and of unit mass. The second relation is between those of
driven oscillator and the undriven oscillator. 
Provided the results in equation (13) are given, these relations give a simple 
method of finding the exact quantum states for a driven harmonic oscillator
system \cite{Kim} or a general quadratic system \cite{Song}, 
as explicitly shown with examples. But a point that should be mentioned 
is that the unitary relation 
method can not give the results in equation (13).

The operator for the first relation is unique up to trivial phase 
\cite{Mosta}, but the other operator which depends on classical solution  
is not unique.

Since the operator of the second transformation is a exponential of
a linear combination of $x$ and $p$, the transformation does not change 
the uncertainties of $x$ and $p$: To be precise, with the quantum states 
of $|n;F>, |n>$ defined as $\psi_n^F = <x|n;F>,\psi_n = <x|n>$, from equation 
(35) one can easily prove the relations
\begin{eqnarray}
&&<n;F|(x-<n;F|x|n;F>)^2|n;F>=  <n|(x-<n|x|n>)^2|n>,\\
&&<n;F|(p-<n;F|p|n;F>)^2|n;F>=  <n|(p-<n|p|n>)^2|n>.
\end{eqnarray}

As a final remark, we add a speculation that there might be some relations 
between a harmonic oscillator system of unit mass time-dependent frequency,  
and a simple harmonic  oscillator.
Independently from the time-dependent Hamiltonian system, Gaussian 
pure states are constructed in \cite{Yuen,Sch} in the study of coherent states.
The $n=0$ wave functions of all time-dependent harmonic oscillator
system belong to those of Gaussian pure states \cite{Song}. 
Our speculation is from the suggestion that the annihilation operator 
of any Gaussian pure state may be obtained from the operator which 
annihilate the ground state of a simple harmonic oscillator \cite{Sch}.

\end{document}